\documentclass[10pt,letterpaper]{article}
\usepackage[top=0.85in,left=2.75in,footskip=0.75in,marginparwidth=2in]{geometry}

\usepackage[utf8]{inputenc}

\usepackage{cite}

\usepackage{nameref,hyperref}

\usepackage{microtype}
\DisableLigatures[f]{encoding = *, family = * }

\raggedright
\usepackage{changepage}

\usepackage[aboveskip=1pt,labelfont=bf,labelsep=period,singlelinecheck=off]{caption}

\newcommand{\revY}[1]{\textcolor{black}{#1}} 

\makeatletter
\renewcommand{\@biblabel}[1]{\quad#1.}
\makeatother

\usepackage{lastpage,fancyhdr,graphicx}
\usepackage{epstopdf}
\fancyheadoffset[L]{2.25in}
\fancyfootoffset[L]{2.25in}

\usepackage{color}

\definecolor{Gray}{gray}{.25}

\usepackage{graphicx}

\usepackage{sidecap}

\usepackage{wrapfig}
\usepackage[pscoord]{eso-pic}
\usepackage[fulladjust]{marginnote}
\reversemarginpar

\begin{document}
\vspace*{0.35in}

\begin{flushleft}
{\Large
\textbf\newline{Magnetophoretic synchronous control of water droplets in bulk ferrofluid}
}
\newline
\\
Georgios Katsikis\textsuperscript{1},
Alexandre Breant\textsuperscript{2},
Anatoly Rinberg \textsuperscript{3},
Manu Prakash\textsuperscript{4,*}
\\
\bigskip
\bf{1} Department of Mechanical Engineering, Stanford University, 450 Serra Mall, California, 94305, USA
\\
\bf{2} Ecole Polytechnique, Route de Saclay, 91128 Palaiseau, France 
\\
\bf{3} School of Engineering and Applied Sciences, Harvard University, Cambridge, Massachusetts  02138, USA
\\
\bf{4} 
Department of Bioengineering, Stanford University, 450 Serra Mall, California, 94305, USA Country, E-mail: manup@stanford.edu
\\
\bigskip
* correseponding@author.mail

\end{flushleft}

\section*{Abstract}
We present a microfluidic platform for two-dimensional manipulation of water droplets immersed in bulk oil-based ferrofluid. Although non-magnetic, the droplets are exclusively controlled by magnetic fields, without any pressure-driven flow.  The diphasic fluid layer is trapped in a submillimeter Hele-Shaw chamber that includes permalloy tracks on its substrate. An in-plane rotating magnetic field magnetizes the permalloy tracks, thus producing local magnetic gradients, while an orthogonal magnetic field magnetizes the bulk ferrofluid. To minimize the magnetostatic energy of the system, droplets are attracted towards the locations of the tracks where ferrofluid is repelled. Using this technique, we demonstrate synchronous generation and propagation of water droplets, analyze PIV data of the bulk ferrofluid flow and study the kinematics of propagation. In addition, we show controlled break-up of droplets and droplet-to-droplet interactions. Finally, we discuss future applications owing to the potential biocompatibility of the droplets.


\section{Introduction}

Magnetic forces can be used actuate objects that are solid (particles), liquid (droplets), or gas (air bubbles) offering potential technological advantages over electric forces\cite{pamme2006:ma}. There is a plethora of techniques for the manipulation of paramagnetic objects such as magnetic microbeads attracted by permanent magnets, electromagnets\cite{deng2001:ma,smistrup2006:mm} or spin-valves \cite{lagae2002:ch}, including beads used in applications towards labeling or separating targeted particles or cells\cite{pammemanz2004:cp,sahoo2005aqueous, pamme2006:co,donolato2011:ma,lim2014:ma}, even blood cells or magnetotactic bacteria \cite{schuler1999:ba,zborowski2003:re,han2004:co,lee2004:co}. Magnetic manipulation has also been applied to weakly diamagnetic objects - examples include the levitation of water femtodroplets\cite{lyuksyutov2004:cp} and small animals, for example frogs in air \cite{simon2000:di} under strong magnetic field gradients. 

While the manipulation of a given object for the above mentioned examples is based on its own magnetic properties, manipulation is also feasible for an object that is non-magnetic, yet immersed in a magnetic medium; the physics of propagation and the interactions of non-magnetic solid objects (beads) immersed in magnetic media, also termed `magnetic holes' have been well studied\cite{toussaint2004:in,de1999:li,de2000:in,rosensweig2013:fe,pieranski1996:br, helgesen1990:dy,vcernak2004:ag,helgesen1990:nl}. In the same category but more on the applied side, many organic entities like cells and large molecules have been assembled in fluid dispersions of magnetic nanoparticles\cite{yellen2005:ar,torres2014recent}. In addition, magnetic holes in the form of water droplets immersed in bulk ferrofluid have been actuated by magnets \cite{zhang2011demand} or miniature coils \cite{padovani2016:electropermanent}  inside pressure-based microchannels. Manipulating water droplets in magnetic media instead of magnetic droplets (ferrofluidic or with some other magnetic trace) in non-magnetic media \cite{schneider2008automated, zhang2009chip,kahkeshani2016drop,varma2016droplet,ray2017demand,sung2017magnetophoretic} may help overcoming biocompatibility limits in the later case, notably the opacity of the ferrofluid itself as well as additional considerations like the pH value, the type of nanoparticles and the stabilizing surfactant and general chemical compatibility \cite{kose2009:la,krebs2009:bc}. However, a purely magnetophoretic method for water droplets in magnetic media, that is without pressure-based channels, is currently missing, thus  rendering simultaneous and precise control of the droplets challenging.

Here we present a platform for magnetophoretic control of water droplets, that is diamagnetic with susceptibility ${\chi}_{p}\sim{-10}^{-5}$ (practically non-magnetic) in bulk hydrocarbon oil-based ferrofluid, that is paramagnetic with initial susceptibility ${\chi}_{m}\sim3$. Thus our system has a negative magnetic susceptibility difference $\Delta \chi={\chi}_{p}-{\chi}_{m}<0$. In our platform, not only the water droplets provide a means of a possible biocompatible encapsulation but also the ferric medium, due to the absolute value $|\Delta \chi|$, increases the magnetic force that is exerted on the water droplets \cite{pamme2006:co} as opposed to utilizing only the diamagnetic nature of water to manipulate droplets with external magnetic fields\cite{lyuksyutov2004:cp}.

\begin{figure*}[t]
\centering
\includegraphics[height=9cm]{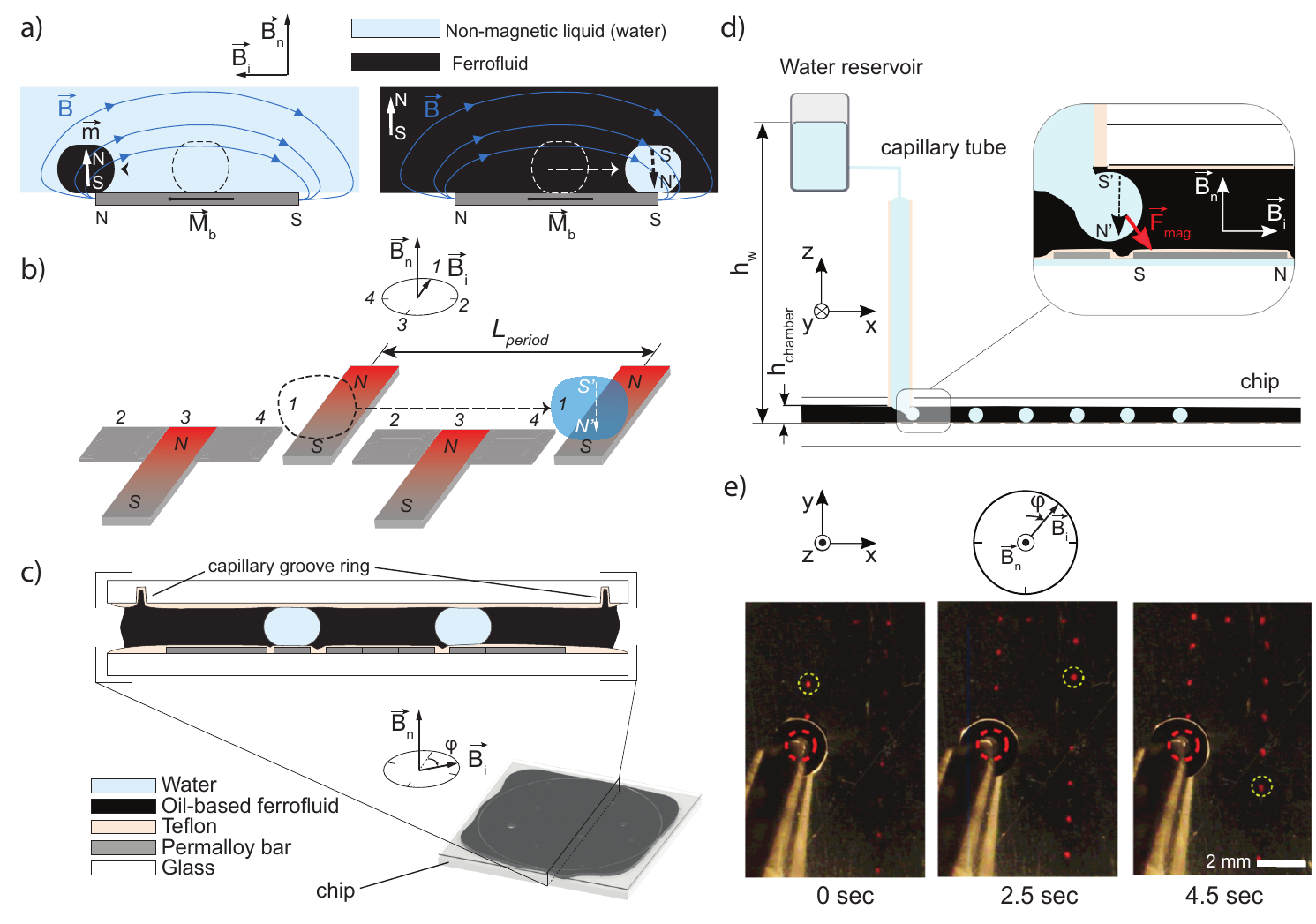}
\caption{Schematic of operating principle and experimental device: (a) Schematic of magnetophoretic approach for manipulation of an object using local gradients of magnetic field ${\bf B}$ generated by a permalloy bar with magnetization ${\bf M}_{b}$  induced by a field  ${\bf B}_{i}$. On the left panel, a ferrofluid droplet inside a bulk non-magnetic liquid is polarized by a field ${\bf B}_{n}$  and moves towards positive field gradients of ${\bf B}$. On the right panel, the bulk fluid is polarized by the field ${\bf B}_{n}$  which is equivalent to the water droplet having an opposite magnetization\cite{rosensweig2013:fe} and thus the droplet moves towards negative field gradients of ${\bf B}$. (b) Permalloy `I' and `T' bars that get magnetized during the rotation of the magnetic field  ${\bf B}_{i}$ , forming a series of poles along the sequence $1-2-3-4$ that creates a propagation pathway for the water droplet (has equivalent polarization with orientation opposite to what  ${\bf B}_{n}$  would induce to a magnetic object). c) Cross-section view of the fluidic chip that contains a capillary groove ring to contain the instabilities of the bulk oil-based ferrofluid. \revY{d) Schematic of water droplet generator in bulk ferrofluid which is a modified set-up from previous work\cite{rinberg2017dropgen}. The water is supplied through a teflon tube at a height ${h}_{w}$ from the surface of the chip. The height of the chamber is symbolized with ${h}_{chamber}$. The magnetic force ${\bf{{F}_{{mag}}}}$ generating the droplet from the teflon tube is shown with red color. The color notation is the same as in (c).  e) Sequential snapshots from experiments with the set-up of (d) generating droplets propagating on winding tracks of `I' and `T' bars with $f=2~Hz$, $B_{i}=40~G$, $B_{n}=250~G$.  The structure with the mustard color is the teflon tube. The droplets appear with orange-red color because of presence of ferrofluidic film on the top glass. The red dashed circles indicate the outlet of the teflon tube and the yellow dashed circle the position of a given droplet at $t=0, 2.5, 4.5~sec$.}}
%
%
\label{Figure_1}
\end{figure*}	

The magnetophoretic control in our platform is based on using external magnetic fields to magnetize permalloy tracks that create propagation pathways for the droplets. A similar approach has been implemented with water-based ferrofluid droplets immersed in silicone oil\cite{katsikis2015:sd} where $\Delta \chi>0$. Here, in comparison to that previous implementation, we take the inverse combination of fluids; the droplets are water (non-magnetic) and the surrounding liquid is magnetic. In particular, the surrounding liquid gets magnetized by two magnetic fields. The first magnetic field ${\bf B}_n$ is perpendicular to the planar geometry, has a fixed magnitude  and uniformly magnetizes the surrounding liquid. The second magnetic field  ${\bf B}_i$ lies on the planar geometry, is rotating and magnetizes soft magnetic permalloy bars that generate a non-uniform magnetic field ${\bf B}$. Under this configuration, it has been proven mathematically\cite{rosensweig2013:fe} that the water droplet, although non-magnetic, has an equivalent polarization that is the opposite to that of the surrounding liquid. Intuitively, the permanent vertical magnetic field is responsible of a "magnetic pressure" on the water droplet, associated with the magnetization drop across the ferrofluid-water interface. Consequently, the magnetic field gradient generated by the bars induce a magnetization difference at the front and back of the droplet, resulting in a pressure difference which sets the droplet into motion. Therefore,for $\Delta \chi\sim-3$, the water droplets are attracted towards the negative gradients of ${\bf B}$ as opposed to a ferrofluid droplet in non-magnetic liquid which would be attracted towards positive gradients (Fig. \ref{Figure_1}a). Utilizing this attraction for the water droplets, we used permalloy patterns of `T' and `I' bars that get magnetized during the rotation of the field ${\bf B}_i$, creating propagation pathways for the droplets.

\section{Experimental section}

\subsection*{Scaling analysis}

In our system, the droplets are of sub-millimeter size (diameters $d=~200-800~mm$) and propagate inside a Hele-Shaw flow chamber that consists of a bottom glass slide that supports permalloy bars (typical lengths $l_{bar}\approx 1~mm$) and a top glass slide that restrains the bulk ferrofluid and prevents evaporation. The height of the flow chamber is $h_{chamber}=400~mm$, resulting in $h_{chamber}<d$, meaning that the droplets are generally tightly confined between the top and bottom slides. 

To gain some insight into the  physics of our system, we wrote four scaling  numbers. First, we calculated the  capillary number $Ca$ based on the dynamic viscosity of the oil ${\mu}_{oil}$, the typical velocity of the droplets ${V}_{drop}$ and the interfacial tension $\sigma$. We defined the velocity scale as ${V}_{drop}=f{L}_{period}$, where $f$ is the frequency of the rotating field ${\bf B}_{i}$ and ${L}_{period}$ is the length of the periodic block of `T' and `I' bar path that defines the travelled distance within a full rotation of ${\bf B}_{i}$ (Fig. \ref{Figure_1}b). Then we wrote the capillary number as:

\begin{equation}
{Ca}=\frac{{\mu}_{oil}f{L}_{period}}{\sigma}
\label{Capillary_number}
\end{equation}

The  capillary number expresses the ratio between the  viscous and  capillary  forces in our system . Using  typical values of  ${\mu}_{oil}=5~mPas$, ${L}_{period}=1.5~mm$, $f=5~Hz$, (Fig.\ref{Figure_2}d) and $\sigma={4}\cdot{10}^{-3}~N/m$, we calculated  $Ca\approx{10}^{-2}$ which means that the droplets are not expected to break-up due to viscous forces, being held together by the capillary forces which are greater than the former by at least an order of magnitude. 

Next, to gain insight into the break-up of droplets due to magnetic forces, we calculated the magnetic Webber number ${We}_{m}$ as the ratio between the magnetostatic energy ${E}_{m}$ which tends to increase the surface area of the bulk ferrofluid under the influence of the magnetization ${\bf M}_{n}$  (in other words,  break the bulk ferrofluid into distinct droplets) over the surface tension energy ${E}_{\sigma}$ that tends to minimize the total surface area thereby preventing break-up. We defined the energy ${E}_{m}$ and based on the fact that the droplets are tightly confined within the flow chamber ($h_{chamber}<d$) we approximated the shape of the droplet as a rigid cylinder of diameter $d$ and height $h$. We therefore defined the volume ${V}_{cylinder}=\pi{d}^2{h}/4$, and write ${E}_{m}={\mu}_{o}{{M}_{n}}^{2}{V}_{cylinder}/2$ where ${\mu}_{o}$ is the magnetic permeability of the vacuum. In addition, we  defined the energy ${E}_{\sigma}=\sigma{S}_{cylinder}$ where ${S}_{cylinder}=\pi d h+\pi {d}^2/2$ is the surface of the cylinder. By assuming $h\sim d$ and combining the expressions for ${E}_{m}$ and ${E}_{\sigma}$, we obtained the magnetic Webber number:

\begin{equation}
{We}_{m}=\frac{{\mu}_{o}{{M}_{n}}^{2} d}{12\sigma}
\label{Mag_Webber_number}
\end{equation}

The magnetic Webber number indicates that a stable droplet volume can be achieved if equation (\ref{Mag_Webber_number}) is unit one. To get the maximum droplet size that is stable,  we set equation (\ref{Mag_Webber_number}) equal to one and solved with respect to the droplet diameter $d$ and obtained
${d}_{max,stable}=12\sigma/({\mu}_{o}{M}_{n}^{2})$. In our system, for the order of magnitude  ${M}_{n}=250~G$, and ${\mu}_{o}=4\pi \cdot {10}^{-7} N/{A}^{2}$, we calculated ${d}_{max,stable}\sim 100~ \mu m$ which indicates the instability of the bulk ferrofluid volume. However, the fact that ${d}_{max,stable}<d$ in our system means that we can use magnetic forces to break-up droplets  at will from the bulk ferrofluid-water interface.

In addition, we calculated the Reynolds number $Re$ based on ${\mu}_{oil}$, the density of the oil ${\rho}_{oil}$, ${V}_{drop}=f{L}_{period}$ and the droplet diameter $d$:

\begin{equation}
{Re}=\frac{{\rho}_{oil}f{L}_{period}{d}}{{\mu}_{oil}}
\label{Reynolds_number}
\end{equation}

The  Reynolds number expresses the ratio between the  inertial and  viscous forces in the flow region around the droplets. For the values we used to evaluate equation (\ref{Capillary_number}) and ${\rho}_{oil}=1.28kg/{m}^{3}$, $d=400~mm$, we calculated  $Re\approx{1}$ which indicates that inertial forces in the flow region around the droplets is expected to be comparable to the  viscous forces.  However, since  we used the droplet  diameter $d$ as the relevant length scale for the Reynolds number in equation (\ref{Reynolds_number}), our argument for the balance between the inertial and the  viscous forces concerns the  bulk region around the droplet and not the thin lubrication  film in the contact region between the droplets the the top and bottom plates where the relevant  length scale is the thickness of the lubrication film $e$ which is  challenging to calculate based on the complex rheology of the ferrofluid, the  magnetic fields affecting the lubrication film and the surface energies of the ferrofluid-water-plate interface. Nevertheless, we expect that the lubrication film should be orders of magnitude smaller than the droplet diameter $e<<d$, from previous works \cite{maruvada1996:retard, rabaud2011:manipul,ling2016:drop,zhu2016:pancake} where bubbles or droplets are tightly confined  in the Hele-Shaw chamber similar to our system ($h_{chamber}<d$). Even though, the $e$ is not a tractable measure like $d$, we expect that the flow in the contact area between the droplet and the plates to be dominated by viscous forces rather than inertial forces.

Last, we calculated the Stokes number $St$ as the ratio between the viscous relaxation time scale ${\tau}_{relax}$ over the forcing time scale ${\tau}_{forcing}$. We defined the viscous relaxation time scale as ${\tau}_{relax}\approx{\rho}_{oil}{d}^{2}/{\mu}_{oil}$ expressing the relevant time scale for a particle with diameter ${d}$ moving at a fluid medium with density ${\rho}_{oil}$ and viscosity ${\mu}_{oil}$ to come to a halt based on the effect of viscous forces. Note that this formula of ${\tau}_{relax}$ represents an upper limit for two reasons. First, we used the droplet diameter $d$ as the length scale even though most of the viscous dissipation is expected to occur at the contact area between the droplet and the plates \cite{rabaud2011:manipul,ling2016:drop}, where the relevant length scale is the thickness of lubrication film $e$ for which $e<<d$. Second, in the factors greater than one that results in the derivation \cite{brennen2005:fund} of the formula for ${\tau}_{relax}$. Furthermore, we defined the forcing time scale as ${\tau}_{forcing}\approx 1/f$, where $f$ is the frequency of the rotating magnetic field ${\bf B}_{i}$ which periodically sets the droplets into motion. Combining these two time scales we can write:

\begin{equation}
{St}=\frac{{\rho}_{oil}{f}{d}^2}{{\mu}_{oil}}
\label{Stokes_number}
\end{equation}

The Stokes number expresses how fast the droplet can react to a disturbance (in our case the periodic forcing by ${\bf B}_{i}$ with frequency $f$) with respect to the viscous forces which determine how the fluid can transmit the momentum disturbance, in other words dissipate momentum. Here, for the same values used in equations (\ref{Capillary_number},\ref{Reynolds_number},\ref{Stokes_number}), we calculated  $St={10}^{-1}$ which as an upper limit of low value, indicates that the droplet reacts fact to the periodic forcing $f$, dissipating momentum instantaneously. Therefore, we can approximate the motion of the droplet as quasi-steady where at any given time instance the magnetic force  ${\bf F}_{m}$ is matched with the hydrodynamic force ${\bf F}_{d}$.

In addition to writing down scaling numbers for the moving water droplets, we made scaling arguments for particles - smaller than the water droplets - that we seeded into the bulk ferrofluid to perform Particle Image Velocimetry (PIV) analysis for visualizing the bulk flow. The movement of the seeding particles should effectively visualize the flow of the bulk ferrofluid if two conditions are met: first, their inertial forces are negligible allowing them to respond immediately to any disturbance from the bulk flow and second, their magnetic forces are much smaller than  the hydrodynamic resistance forces. In terms of the first condition, we invoked equation (\ref{Stokes_number}) and calculated  $St\approx~{10}^{-5}$ for typical size of seeding particles\cite{westerweel1997:fundamentals} of diameter ${d}_{p}=10~\mu m$  thus showing that this condition is met. In terms of the second condition, the magnetic forces on the seeding particles are non-zero because the particles do not have the same magnetic susceptibility coefficient with the bulk ferrofluid, that is ${\chi}_{p}={\chi}_{m}$ in equation (\ref{Force_with_xp_xm_difference}). Thus the seeding particles, similar to the water droplets, behave as `magnetic holes'. To estimate the how the magnetic forces compare to the hydrodynamic forces induced by the motion of the surrounding bulk fluid, we wrote down scaling arguments for their magnitudes.

The magnitude for the magnetic forces  on the particle can be written as  ${F}_{mp}=V_{p}|\vec{\bigtriangledown}({\bf M}_{p}{\bf B})$| where  $V_{p}=4/3\pi{{d}_{p}}^2$ is the volume of the spherical particle, ${\bf M}_{p}=-{\bf M}_{n}$ is the equivalent magnetization of the  particle that is  equal and opposite to that of the bulk ferrofluid\cite{rosensweig2013:fe}, and ${\bf B}$ is the magnetic field generated by the bar. Assuming  uniform magnetization ${\bf M}_{p}$ and a gradient of magnetic field  ${\bf B}$ along a single reference axis, $x$ that is $|\vec{\bigtriangledown}{\bf B}|=\vartheta{B}/\vartheta{x}$, we can write ${F}_{mp}=(4/3)\pi{{d}_{p}}^3 {M}_{n}\vartheta{B}/\vartheta{x}$. The magnitude for hydrodynamic forces on the particle can be written based on Stokes flow\cite{white2006:vi} as ${F}_{dp}=3\pi{\mu}_{oil}{d}_{p}{V}_{p}$ where ${V}_{p}=f{L}_{period}$ is the velocity induced by the  movement of the droplet.  Taking the ratio of the two expressions for ${F}_{mp}$ and ${F}_{dp}$ we get:

\begin{equation}
\frac{{F}_{mp}}{{F}_{dp}}=\frac{4{{d}_{p}}^2{M}_{n}}{9{\mu}_{oil}f{L}_{period}}\frac{\vartheta{B}}{\vartheta{x}}
\label{Ratio_Mag_to_Hydro_Forces}
\end{equation}

For the same values we used for evaluating equations (\ref{Mag_Webber_number},\ref{Reynolds_number}), and $\vartheta{B}/\vartheta{x}\approx 120~G/mm$ from previous modeling work \cite{katsikis2015:sd}, we calculated ${F}_{mp}/{F}_{dp}\approx~{10}^{-10}$. This value shows the dominance of the hydrodynamic over the magnetic forces and  implies that the motion of the particles owning to the magnetic forces will be suspended owning to the motion of the bulk ferrofluid in the proximity with a moving water droplet. Therefore, in our system we can effectively  visualize the  flow of bulk ferrofluid.

\subsection*{Flow chamber}

Both the bottom and top glass slides were made by cutting microscope glass slides of thickness $1~mm$ (Fisherbrand) in $25 \times 25~mm$ squares using a glass cutter with a carbide wheel (Fletcher-Terry company). The patterning of the permalloy bars on the bottom glass slide was carried out using a 3W diode pumped solid state UV laser cutter (DPSS Lasers Inc.). A permalloy foil of thickness $t=25~\mu m$ (Supermalloy C, composition Ni77/Fe14/Cu 5/Mo 4, Goodfellows) was bonded on the bottom glass slide, using Epoxy glue (Loctite E-30CL) and cured at room temperature for 24 hours. Then, to pattern the shapes of the permalloy bars, the laser cutter was programmed to etch the foil. Next, the permalloy foil was peeled off the glass slide, leaving the etched shapes of the permalloy bar shapes. The sample was then cleaned with compressed air and spin-coated with Teflon at 1500 revolutions per minute (rpm) for 30 sec. The Teflon-coated sample was cured on a hot plate at ${90}^{o}C$ for 1 hour. To complete the Hele-Shaw flow chamber, the top glass slide, similarly coated with Teflon,  was added on top of the chip using 0.006'' thick rubber sheets (Mac Master Car) as spacers. The top glass slide has a carved circle of depth of $20~ \mu m$ etched using the laser cutter. The carved circle acted as a capillary groove ring (Fig. \ref{Figure_1}c) that prevented the bulk ferrofluid from forming instabilities causing leakage out from the flow chamber ${\bf B}_{n}$ as indicated by evaluating ${d}_{max,stable}\approx 100\mu m$ .

\subsection*{Imaging system}

\begin{figure*}[t!]
\centering
\includegraphics[width=13cm]{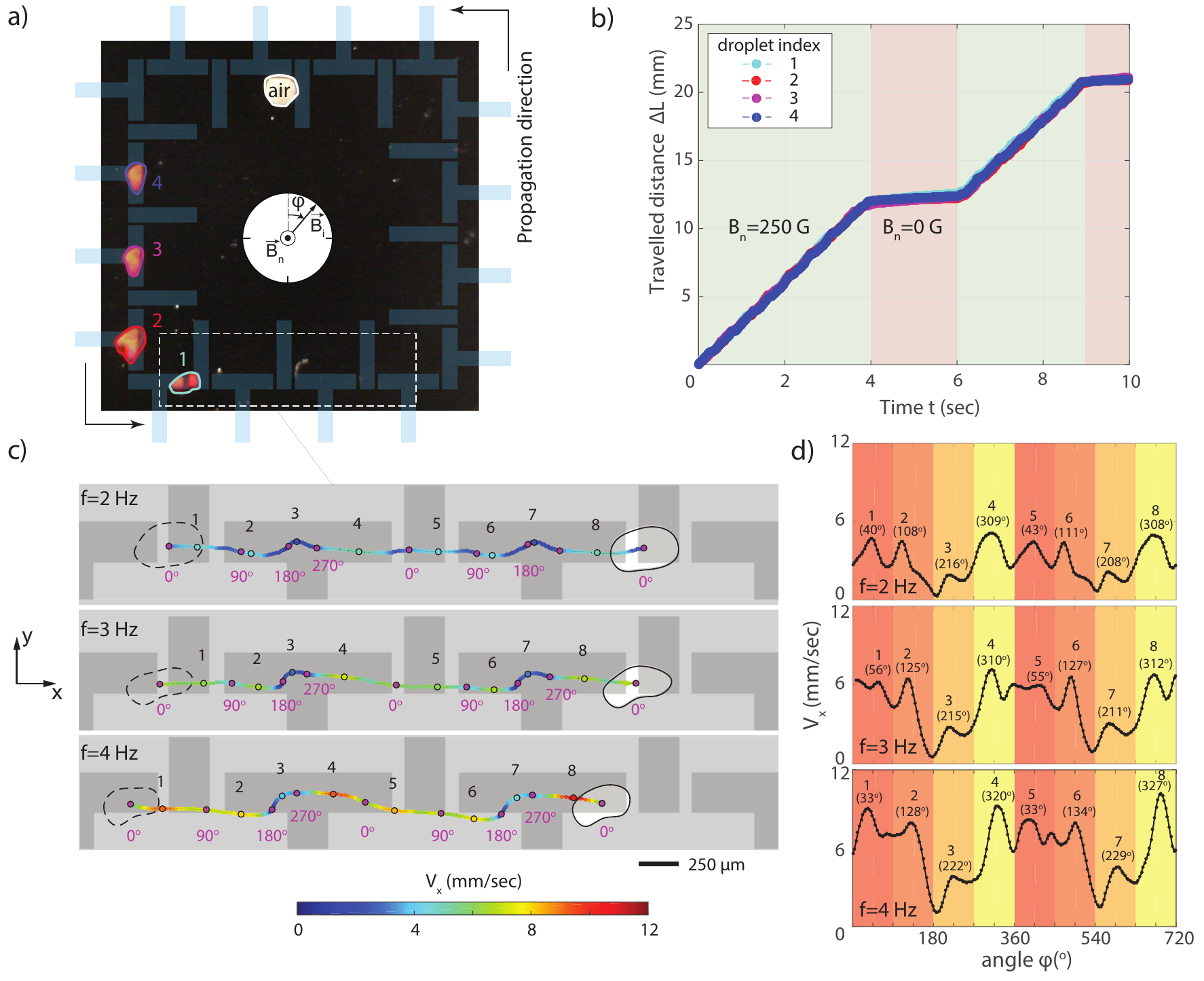}
\caption{ Kinematics of propagating water droplets, immersed in bulk oil-based ferrofluid. (a) Snapshot of four droplets, noted by numbers $1-4$, propagating in a counter-clockwise direction on a closed loop of `T' and `I' bars, drawn with light blue color. b) Kymogram of travelled distance $\Delta L$ of the droplets of (a) versus time $t$. The green and red color-bands indicate bias field $B_n=250~G$ and $B_n=0~G$ respectively. c) Droplet trajectories of a water droplet of diameter $d=300~\mu m$  (channel height  $h=380~\mu m$) where the in-plane rotating field ${\bf B}_{i}$  is rotating a clockwise at three frequencies, $f=2,3,4~ Hz$.  The color of the trajectories represents the magnitude of the horizontal component of ${V}_{x}$. The numbers $1-8$ denote the local velocity maxima ${V}_{x}$. Purple numbers/circles indicate the four cardinal angles $\varphi=(0^{o},90^{o},180^{o},270^{o})$. The trajectories are repeatable over each ${360}^{o}$ cycle. The image background has been removed from the images and the grey color indicates the bulk oil-based ferrofluid.  The angle $\varphi$ denotes the instantaneous angular orientation of ${\bf B}_{i}$ (a) and is proportional to normalized time $tf$ with $\varphi(t)=2\pi ft$. The parentheses indicate the angles $\varphi$ that correspond to each potential well. d) Diagrams of Velocity $V_{x}$ versus  $\varphi$ for $f=2-4~Hz$. Black numbers indicate the same $V_{x}$ maxima shown in (a) and the numbers in the parentheses indicate the angles $\varphi$ where these maxima occur. The vertical color-bands indicate each quadrant of $\varphi$. Experiments performed at  $B_{i}=40~G$, $B_{n}=250~G$. Viscosities ${\mu}_{drop}=1~\mu Pas$, ${\mu}_{oil}=3~\mu Pas$.}
\label{Figure_2}
\end{figure*}

To obtain high-speed video data of the propagating water droplets, a custom darkfield illumination imaging system was developed. A brightfield illumination imaging system utilizing reflected microscopy had been used before\cite{katsikis2015:sd} but could not attain sufficient contrast to image the water droplets because of the opacity of the bulk ferrofluid oil. In order to enhance contrast, a custom illumination ring was created by machining a board plate with copper coating of $\approx 0.5 ~mm$  using a programmable micromil (Modela RDX-20, Roland) to create copper pads for the serial connection of ten high power LEDs (White 4000K, Mouser electronics). The LED ring was driven by a DC power supply (GW Instek GPD4303S) at a voltage of approximately 30V. This LED ring was placed on top of the chip (Fig. \ref{Figure_2}d) and mounted on a microscope objective (Mitutoyo Infinity-Corrected lens 2X, 10X). The optical path from the objective leads to a high-speed camera (Phantom Camera v1210 Vision Research) through a ${90}^{o}$ mirror. The sampling frequency ${f}_{s}$ during the image acquisition was in 30 to 2000 frames per second. The darkfield illumination through the peripheral lighting provided by the LED ring allowed for high contrast in the video images, revealing a definite interface between the water droplets (high brightness region) and the bulk ferrofluid (low brightness region).

\subsection*{Apparatus of electromagnets}

The two magnetic fields ${\bf B}_{i}$, ${\bf B}_{n}$ were induced to the microfluidic chip through a system of electromagnetic coils\cite{katsikis2015:sd}. The same system included a cooling system for the chip, a microcontroller, and Hall effect sensors to measure the magnitudes of the magnetic field.

\subsection*{Fluids}

For our working fluids, an immiscible solution of deionized (DI) water droplets in light hydrocarbon ferrofluidic oil (EMG 905, Ferrotec) was used. The viscosity of the ferrofluidic oil was ${\mu}_{f}=3~cSt$. To insert the fluid in the flow chamber, the ferrofluid was first pipetted on the chip (volumes ${V}_{f}=100-140~\mu L$) . Then, the water was added, with a surfactant volume concentration  ${C}_{T20}=0.06-0.6\% $ (Tween 20, Fisher Scientific). The surfactant, by decreasing the interfacial tension, enables the droplets to further deform and thus to increase the magnitude of the magnetization difference accross its length, resulting in an stronger actuation force. In addition, the surfactant is qualitatively expected  that the decrease in interfacial tension will reduce the hydrodynamic drag exerted on the moving droplet \cite{dangla2012:2d}.

\subsection*{Image Analysis}

In order to track the droplets in the raw video data, a custom code was written in MATLAB. Owing to the darkfield illumination system, the droplets appeared as high brightness pixel regions. The droplets were then tracked based on a brightness threshold set on the each frame of the video sequence after subtracting the average image of the sequence. Multiple droplets were tracked using a nearest-neighbor tracking algorithm. The centroid coordinates as well as other geometric dimensions were extracted out of those regions, after eroding and dilating the pixel regions representing the droplets. The coordinates of the centroids were smoothed using a moving average filter with a span value equal to five. Then, the velocities of the droplets were calculated by using the second order finite-difference scheme on the smoothed coordinates of the droplets over subsequent frames. The boundaries of the droplets were derived by interpolating splines on the edges of the pixel regions representing the droplets. The asynchronization events (loop occurrences) were automatically detected by comparing the actual traveled distance of the droplet to the one expected.

\subsection*{Particle Image Velocimetry (PIV)}

To visualize the flow of the bulk ferrofluid, the bulk ferrofluid was seeded with Micro-Fine Teflon (Polytetrafluoroethylene or PTFE) particles (Howard Piano industries)  which had an average diameter of ${d}_{p}=3~\mu m$.  These teflon particles were  selected  because they maintained a distinct optical signal within the dark brown opaque oil-based ferrofluid even though they did not have the same magnetic susceptibility coefficient as the bulk ferrofluid; the Teflon beads are practically non-magnetic objects with  ${\chi}_{p}=0$ and therefore are set in motion in the bulk ferrofluid behaving as `magnetic holes'. However, it was deemed that they can effectively visualize the flow of the bulk ferrofluid based on the numerical result of equation (\ref{Ratio_Mag_to_Hydro_Forces}). To analyze the PIV data, we used an open source software\cite{thielicke2014:pi}. To improve the accuracy of the analysis, the PIV analysis was applied only where the bulk ferrofluid was present; using the image analysis code all water droplets that were tracked and their corresponding regions were excluded from the PIV analysis. 

\section{Results and discussion}

\subsection*{Kinematics of propagating droplets}

We first performed experiments with water droplets immersed in bulk oil-based ferrofluid (Fig. \ref{Figure_1}b,c). Upon activation of the magnetic fields ${\bf B}_{i}$, ${\bf B}_{n}$, the water droplets started propagating on top of the `T' and `I' bars. Our system is capable of manipulating multiple droplets simultaneously (Fig. \ref{Figure_2}a) using these two magnetic fields ${\bf B}_{i}$, ${\bf B}_{n}$. The deactivation of the magnetic field ${\bf B}_{n}$ can cause the droplets to stop their propagation momentarily until ${\bf B}_{n}$ is activated again (Fig. \ref{Figure_2}b and supplementary video 1) thus causing the droplets to resume their propagation. Furthermore, reversing the rotation of ${\bf B}_{i}$ can cause the droplets to switch the direction of their propagation; droplets propagating in closed loops of `T' and `I' bars switched from propagating clockwise to counterclockwise (supplementary video 2). Furthermore, the magnetophoretic control is robust as shown by the overlapping droplet trajectories in the closed loops (supplementary video 2).

	In addition to manipulating multiple droplets simultaneously, we performed experiments with single water droplets in order to study their detailed kinematics. Using our image analysis code, we tracked the propagating droplets and calculated their trajectories and velocities. We observed that the trajectories of the propagating droplets are serpentine (Fig. \ref{Figure_2}a and supplementary video 3) owing to the fact that the poles formed on the bars do not lie on a straight line. The propagation is a result of the rotation of the field ${\bf B}_{i}$ which activates the alternating magnetic poles on the `T' and `I' bars making pathways for the droplets. Therefore, when the field ${\bf B}_{i}$  rotates at a higher frequency, the droplets propagate at higher speeds (Fig. \ref{Figure_2}b). If we compare the droplet trajectories (Fig. \ref{Figure_2}a) for frequency $f=2~Hz$ versus $f=~3Hz$, we observe that for each of the four cardinal angles $\varphi=(0^{o},90^{o},180^{o},270^{o})$ of ${\bf B}_{i}$, the droplet at $f=2~Hz$ has traveled more distance across the `T' and `I' bars than the droplet at $f=3~Hz$. We observe the same phenomenon when comparing the trajectories for $f=3~Hz$ versus $f=4~Hz$. We attribute this phenomenon to the greater viscous dissipation that  occurs as a result of the higher droplet speeds induced at increased frequencies $f$. In other words, as the frequency $f$ increases, it becomes more difficult for the droplet to follow the activated poles on the `T' and `I' bars, resulting in a delay within the cycle of $\bf{{B}_{i}}$ that gets corrected, nevertheless, by the end of the cycle and the propagation remains synchronous. The same phenomenon had previously been observed in a similar magnetophoretic platform with `T' and `I' bars using the inverse combination of fluids, that is, ferrofluid droplets in oil carrier fluid\cite{katsikis2015:sd}. Similar to that system, here the viscous forces dominate over the inertial forces, also indicated by $St\approx 0.1$ from equation (\ref{Stokes_number}) and the droplets propagate with a pattern of sudden accelerations and decelerations (Fig. \ref{Figure_2}d) induced by each of the distinct magnetic poles that get activated the angular orientations of ${\bf B}_{i}$ thus prescribing the droplet trajectories.

\subsection*{Efficiency of propagation}

\begin{figure}[h]
\centering
\includegraphics[height=7cm]
{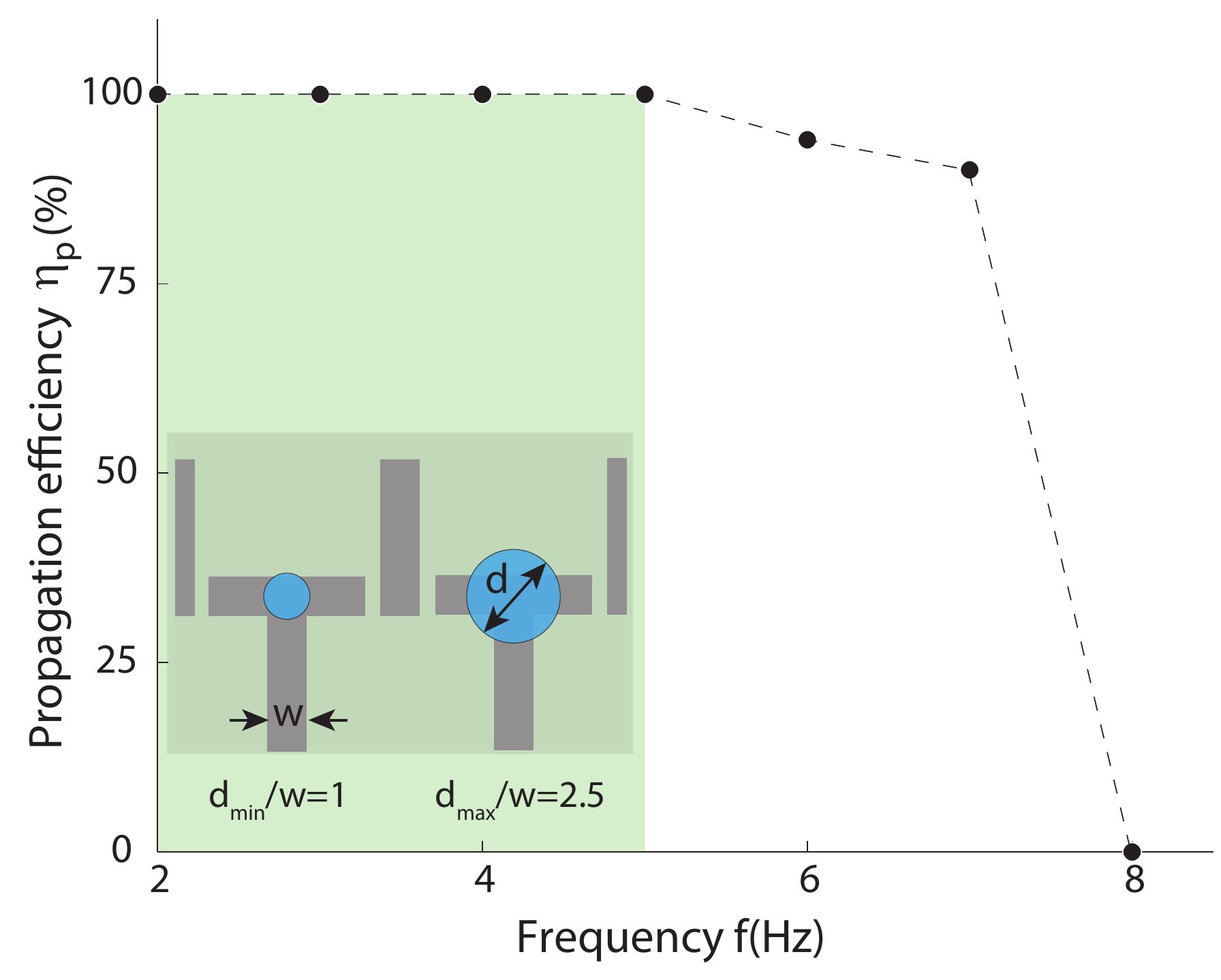}
\caption{Plot of the efficiency of propagation ${\eta}_{p}$ over frequency $f$ for a range of droplet sizes $d=300-1000~\mu m$ and fixed channel height $h=380~\mu m$. The efficiency of propagation ${\eta}_{p}$ is calculated based on equation (\ref{Prop_efficiency}). The regime $f\leq 5~Hz$ with ${\eta}_{p}=100\%$ is highlighted with green color and the inset illustrates the range $[{d}_{min}/w, {d}_{max}/w]$ of droplet diameters $d$ that propagate synchronously in that regime where $w$ is the width of the bars.}
\label{Figure_3}
\end{figure}

In addition to the droplet kinematics (Fig. \ref{Figure_2}), we explored the efficiency of propagation, that is the limit of frequency $f$ after which the droplets will lose their synchronization. To realize this experimentally, we studied the propagation of droplets (diameters $d=250-800~\mu m$, or $1-3$ times the width of the bars) in linear tracks of `T' and `I' bars for frequencies in the range $f=2-8~Hz$. In these ranges, we observed that propagation of the droplets remains synchronous with the rotation of the field ${\bf B}_{i}$ for frequencies $f\leq 5~Hz$ and droplet diameters $d$ from $1$ to $2.5$ times the width $w$ of the bars  (Fig. \ref{Figure_3}). In other words, for $f\leq 5~Hz$, the water droplets propagate covering the length ${L}_{period}$ of the periodic block of T' and `I' bars for every full of rotation of ${\bf B}_{i}$. However, for $f>5~Hz$ the droplets occasionally fail to propagate for a number ${N}_{a}$ out of a total number of rotations $N_t$ of $\bf{B_{i}}$, that is the droplets lose their synchronization with ${\bf B}_{i}$ . To characterize the occurrences of these asynchronization events, we defined the propagation efficiency ${\eta}_{p}$:

\begin{equation}
{\eta}_{p}=\frac{{N}_{t}-{N}_{a}}{{N}_{t}}
\label{Prop_efficiency}
\end{equation}

We observed that for $f>5~Hz$, the efficiency of propagation decreased gradually leading to ${\eta}_{p}=0$ at $f=8~Hz$ (Fig. \ref{Figure_3}). We attribute this failure to the delay within the cycle due to the greater viscous forces, similar to the delay we observed for $f< 5~Hz$ (Fig. \ref{Figure_2}), however for $f> 5~Hz$, this delay cannot be compensated for within the cycle, leading to asynchronization events and, eventually to no propagation (Fig. \ref{Figure_3}) 

Interestingly, we observe that the shape of the droplets in the range $d=300-1000~\mu m$ changes during propagation (supplementary videos 1-2), which we expected since the droplets were over the stability limit ${d}_{max,stable}\approx 100\mu m$ we derived from equation (\ref{Mag_Webber_number}). However, we observed that the droplets did not break-up while propagating in the linear tracks of `T' and `I' bars.

    \revY{In terms of the kinematics and efficiency of propagation, it is interesting to compare the current system to the previous one with the inverse combination of fluids \cite{katsikis2015:sd} in which ferrofluid droplets were manipulated in silicone oil. The kinematics and efficiency as well as shape deformation are determined by interplay between the magnetic force and the hydrodynamic resistance force. From a physical standpoint, we would expect the two systems to be practically identical if the parameters for these forces were matched. In particular, the magnetic force depends on the difference $\Delta \chi={\chi}_{p}-{\chi}_{m}$ between the magnetic susceptibilities of the particle and medium correspondingly. The hydrodynamic drag depends on the interfacial tension $\sigma$ between the two liquid phases and the coated wall surfaces, and the viscosities of the droplets and bulk fluid ${\mu}_{drop}$, ${\mu}_{oil}$. While the kinematics are qualitatively similar between the two systems, exact matching was not possible due to differences in these parameters as well as some technical considerations. For example, in the current system we expected that the teflon layer at the coated surfaces constituted a hydrophobic, yet not oleophobic coating for the bulk ferrofluidic hydrocarbon oil resulting in different wetting properties; in practice we observed that the teflon layer inside the top glass (Fig. \ref{Figure_1}c) was wetted by the bulk ferrofluid and the water droplets appeared with a orange color (Supplemetary Videos 1,3). The teflon layer also showed faster deterioration being exposed to the bulk ferrofluidic hydrocarbon oil than the bulk silicone oil\cite{katsikis2015:sd} affecting the performance, possibly contributing to a lower synchronization limit of $5~Hz$ as opposed to $10~Hz$ in the previous system.}

\subsection*{PIV analysis of bulk ferrofluid}

\begin{figure}[t]
\includegraphics[height=6cm]{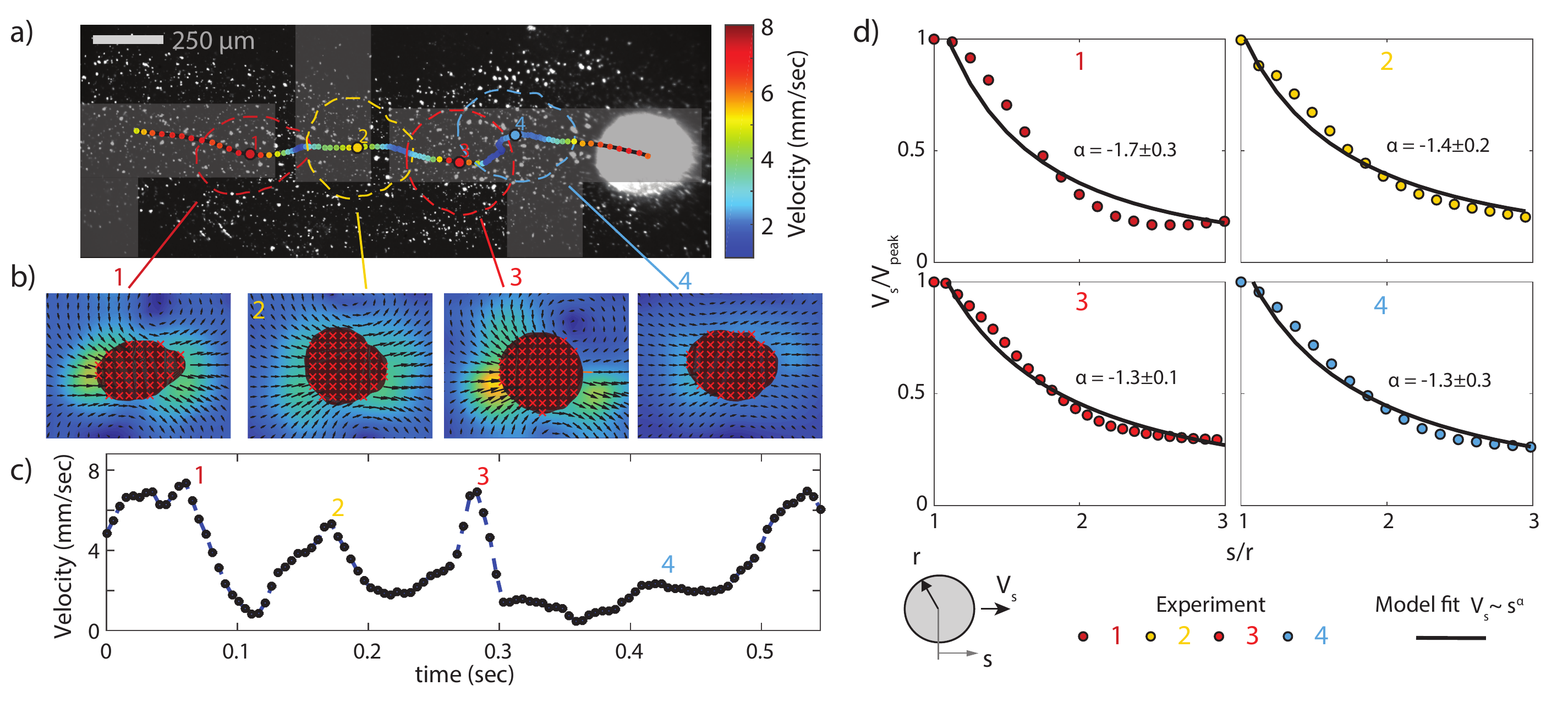}
\caption{PIV analysis of the flow field around droplet of diameter $d=350~\mu m$ propagating at $f=2~Hz$ with $B_{i}=40~G$, $B_{n}=250~G$. a) Snapshot of a single water droplet, shown as a bright spot, propagating rightwards. The smaller bright spots are the seeding particles. The trajectory of the droplet is colored based on the instantaneous magnitude of the velocity ${\bf V}$ of the droplet. The positions, where local velocity peaks ${V_{peak}}$ of ${\bf V}$ occur, are marked with the numbers $1-4$. The corresponding droplet shapes at positions $1-4$ are colored according to the values of ${V_{peak}}$. b) Contours and vector fields of the velocity of the droplet ${\bf V}$ for the positions $1-4$.  c) Magnitude of velocity ${\bf V}$ as a function of time $t$. Each velocity peak ${V_{peak}}$ at positions $1-4$, shown in (a), (b) is also marked. \revY{d) Plots of velocity ${V}_{s}$ of flow field of bulk ferrofluid along the direction $\bf{s}$ of instantaneous droplet velocity ${\bf V}$ at positions $1-4$ with similar coloring scheme as in (c). The velocities ${V}_{s}$ are normalized by ${V_{peak}}$ and the distance $s$ by the radius of the droplet $r$. The black curves are model fits of the form ${V}_{s}\sim{s}^{\alpha}$ where $\alpha$ is calculated with non-linear least-squares method. The estimation ranges, shown with $\pm$, indicate $95\%$ confidence bounds.}}
\label{Figure_4}
\end{figure}

Aside from the kinematics of the water droplets, we studied the movement of the bulk ferrofluid using PIV analysis. In particular, we tested the displacement of the seeding teflon particles in the bulk ferrofluid covering the permalloy bars that were under the influence of the two magnetic fields ${\bf B}_{i}$, ${\bf B}_{n}$. We studied the movement of the bulk ferrofluid both in the absence and in the presence of propagating water droplets (supplementary video 4). 
    In the absence of water droplets, the particles oscillate without translating. This oscillation is the result of the magnetic force that the beads are subjected to as they effectively behave as magnetic holes. However, this magnetic force cannot result into a net propagation because the magnetic force is too low in comparison to the hydrodynamic resistance force that the beads would need to overcome if they were to move over the length scales of the bars, as indicated by the result of equation (\ref{Ratio_Mag_to_Hydro_Forces}).Nevertheless, we noticed that in practice this magnetic force was sufficient to cause the particles to concentrate in the vicinity of the `T' and `I' bars (Fig. \ref{Figure_4}a and Supplementary Video 4).  As expected from continuity of the bulk ferrofluid behaving as an incompressible fluid, the oscillatory motion of the particles shows that the bulk ferrofluid does not translate by itself even though it is subjected to magnetic forces. The only net motion of the ferrofluid that we observed was located at the edge of the chip were there is the interface of the ferrofluid with the surrounding air. 
    
    In the presence of propagating water droplets, the seeding particles translate, thus mapping the flow of the bulk ferrofluid around the propagating droplets (Fig. \ref{Figure_4}b, Supplementary Video 5). From a geometric point of view, our system bears resemblance to Hele-Shaw configurations of confined droplets driven by pressure differences\cite{beatusbarziv2012:ph}\revY{. In these systems, typically with $Re\sim$ ${10}^{-4}$ to ${10}^{-2}$, the inertial forces in the flow field around the droplets are negligible in comparison to the viscous forces. Using the Darcy approximation, the flow field around the droplet is given by a 2D potential in a quasi-2D channel and by taking the two sidewalls as a sum of image dipoles, a tractable mathematical formulation is derived leading to a scaling of ${V}_{flow}\sim{r}^{-2}$ where $r$ is the distance away from the surface of the droplet. This scaling ${V}_{flow}\sim{r}^{-2}$ is valid for $r<W$, where $W$ is defined as twice the distance between the center of the confined droplet from the channel wall, where for a tightly confined system (chamber height ${h}_{chamber}$ such that ${h}_{chamber}<d$ where $d$ is droplet diameter) $W=d$. For $r>W$, ${V}_{flow}$ decays exponentially\cite{beatusbarziv2012:ph} related to the degree of confinement. In our system, the inertial forces of the bulk fluid are not negligible as indicated by $Re\approx~1$  from equation (\ref{Reynolds_number}) and we investigated the decay of the velocity of the flow field around the propagating water droplets. To characterize the flow field of the bulk ferrofluid, we extracted the magnitude of the droplet velocity over time as a given droplet propagated across a periodic block of 'T' and 'I' bars \ref{Figure_4}a). Next, we performed PIV analysis on the flow field (Fig. \ref{Figure_4}b) and separated for the four instances during the cycle where the droplet velocity reached a local peak ${V}_{peak}$ (Fig. \ref{Figure_4}b,c: numbers $1-4$). Then, for every one of the four instances where a ${V}_{peak}$ occurred, we calculated the magnitude ${V}_{s}$ of the flow field away from the droplet, across the direction $s$ of its instantaneous velocity (Fig. \ref{Figure_4}d: numbers $1-4$). Using a non-linear least-squares method, we conducted a model fit of the form ${V}_{s}\sim{s}^{\alpha}$ to the experimental data. We thus obtained $\alpha=-1.7$ to $-1.3$ which lies close to the reference value ${\alpha}_{ref}=-2$. We attribute this deviation of $\alpha$ from ${\alpha}_{ref}$ to inertial effects as $Re\approx~1$ which also create a `memory' in the flow field. Since inertia is present, the flow does not respond immediately to the movement of the droplet as the latter propagate a pattern of sudden accelerations and decelerations (Fig. \ref{Figure_2}d). We hypothesize that the deformation of the droplet due to magnetic forces during propagation might also play an effect in the characteristics of the flow field. We will save a more thorough investigation into the scaling of the flow field for future efforts.}
    
\subsection*{Study of droplet break-up}

\begin{figure}[t!]
\centering
\includegraphics[width=8.9cm]{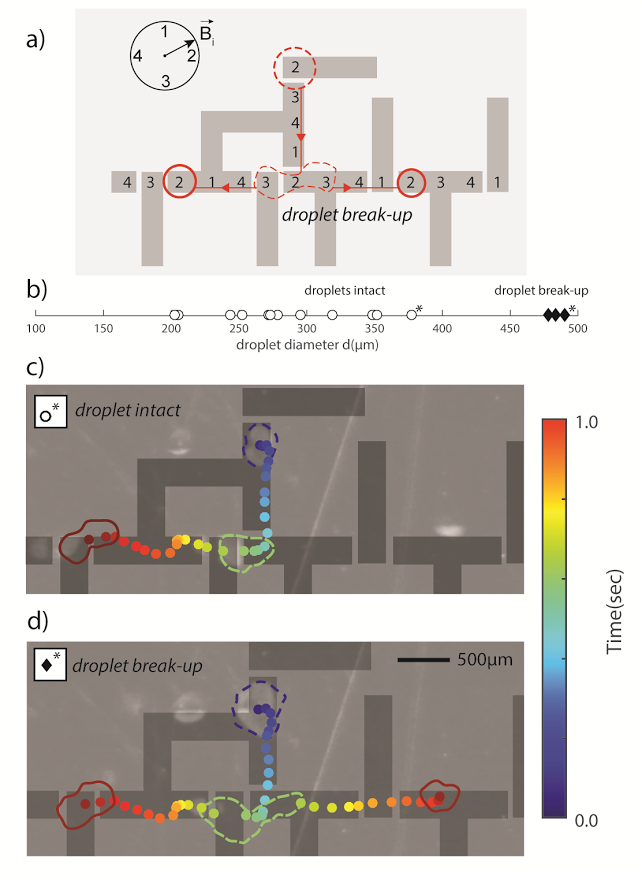}
\caption{Junction of `T' and `I' bars for inducing droplet break-up. a) Schematic of junction showing the break-up of a single droplet (dashed lines) into two satellite droplets (continuous lines) over two cycles of ${\bf B}_{i}$. The black numbers $1-4$ denote the potential wells on the bars activated by the angular orientations of ${\bf B}_{i}$ prescribing the droplet trajectories (red) . b) Graph of the droplet diameters tested where white circles indicate droplets exiting the junction intact. Black diamonds indicate droplets breaking up. c,d) Snapshots from experiments (Supplementary Video 5) where the droplet exits the junction intact (c) or breaks up (d). The asterisks correspond to the points in (b). The snapshots are presented half-opaque and the overlayed dark grey rectangular shapes represent the `T' and `I' bars which are not visible under the opaque bulk ferrofluid. Droplet shapes and trajectories are colored based on the elapsed time. Experiments performed at  $B_{i}=40~G$, $B_{n}=250~G$ and $f=2~Hz$.}
\label{Figure_5}
\end{figure}

The magnetic forces in our system can also be used for functions other than transporting droplets. In fact, we used the same `T' and `I' permalloy bars as basic blocks to create junctions where droplets break-up into satellite droplets (Fig. \ref{Figure_5}a and supplementary video 5). When a droplet arrives at one of these junctions, it is subjected to opposing magnetic forces that pull the droplet along two opposite paths. As a result of these opposing forces, the droplet stretches and divides into two satellite droplets; each of the satellite droplets is diverted into one of the two opposite paths. Previous work has been reported with T-shaped junctions that have been implemented to break droplets in pressure-driven microchannels using shear stress\cite{link2004:ge,garstecki2006:fo,tan2004:de}; however, our design relies solely on magnetic forces that act on the bulk ferrofluid and, in turn, affect the water droplets. 	Here, we created the junction for break-up prompted by 
both the stability limit ${d}_{max,stable}\approx 100\mu m$ that our droplet sizes exceed, and a simple analytical calculation from previous modeling work\cite{katsikis2015:sd} that shows that the magnetic force is comparable to the force owning to the interfacial tension between the bulk ferrofluid and the water. In particular, for the range of droplet diameters of $d=100-1000~\mu m$, we calculated that the magnetic force lies in the range ${F}_{mag}=10-30~\mu N$ and the interfacial tension force that tends to keep the droplet intact is of the order  ${F}_{\sigma}=\sigma \cdot {d}^2=1-10~\mu N$. The magnitudes of ${F}_{mag}$, ${F}_{\sigma}$ are similar showing droplet break-up is possible. 

\begin{figure*}[t!]
\centering
\includegraphics[height=9cm]{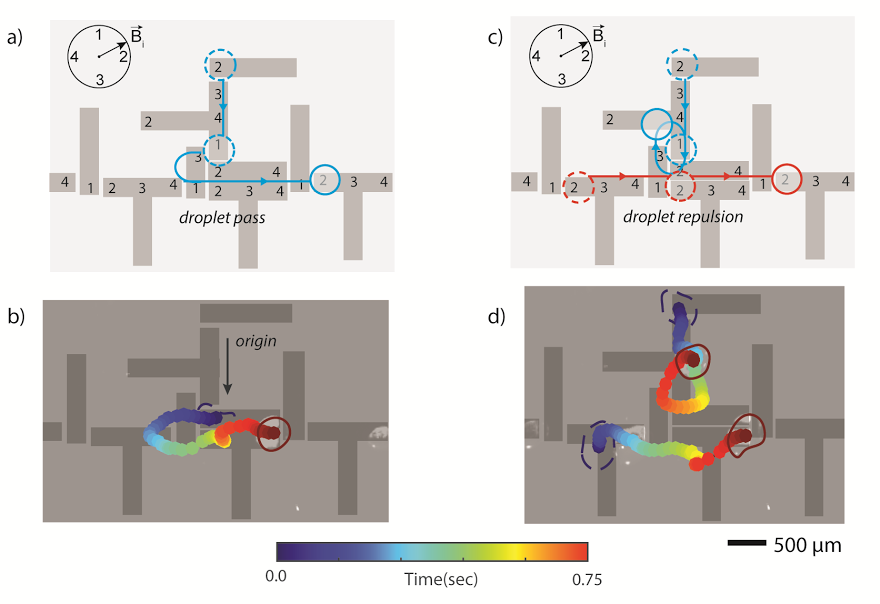}
\caption{Junction of `T' and `I' bars for demonstrating droplet repulsion. a,b)  Schematics of junction for droplet (blue) passing freely on the junction (a) and droplet (blue) repulsed by another droplet (red) crossing the junction (b). The black numbers $1-4$ denote the potential wells on the bars activated by the angular orientations of ${\bf B}_{i}$, prescribing the droplet trajectories, shown in corresponding blue and red colors. The droplets are shown over 3 consecutive cycles ${\bf B}_{i}$ and their shape is shown with sparse dashed, dense dashed and continuous lines. c,d) Snapshots from experiment demonstrating the free passing of droplet (c) and the repulsion of the droplet (d) corresponding to schematics (a) and (b) respectively. As long as a second droplet crosses the junction the droplet coming from the top will be repulsed into a loop over the bars (Supplementary Video 6). The notation is the same as in Fig.\ref{Figure_5}.Experiments performed at  $B_{i}=40~G$, $B_{n}=250~G$ and $f=2~Hz$.}
\label{Figure_6}
\end{figure*}   

	In our system, the droplet break-up takes place as a two-step dynamic process over the course of a half-cycle rotation of the field ${\bf B}_{i}$, that is $\Delta \varphi \sim~180^{o}$. In the first step (${\varphi}\sim~90^{o} \rightarrow 180^{o} $ or bar positions $2\rightarrow3$ in Fig.\ref{Figure_5}a), the droplet gets elongated as a result of two opposing forces exerted by a `I' and a `T' bar placed bilaterally on the droplet. The `I' and a `T' bars are simultaneously activated by ${\bf B}_{i}$ (positions `3'  on Fig.\ref{Figure_5}a). In the second step (${\varphi}\sim~180^{o} \rightarrow 270^{o} $ or bar positions $3\rightarrow4$ in Fig.\ref{Figure_5}a), the elongated droplet is pulled from both sides, forms a neck, and then breaks up as a result of opposing forces from two new activated bars (positions `4'  on Fig. \ref{Figure_5}a). The droplet thus breaks into two satellite droplets that are diverted into two opposite paths. 
	However, we showed experimentally that droplets do not necessarily break up at the junction, unless their diameter is above a threshold size. To characterize this threshold, we investigated the ability of our junction to perform droplet break-up at $f=2~Hz$ for droplet diameters ranging from $d=200-500~\mu m$ (Fig. \ref{Figure_5}b and supplementary Video 5). We found that droplets with diameters smaller than ${d}_{min}\sim 400~\mu m$ do not break-up (Fig. \ref{Figure_5}c). On the other hand, droplets above ${d}_{min}\sim 400~\mu m$ break up into two satellite droplets (Fig. \ref{Figure_5}d) that are not equal in diameter. In fact, the ratio of their diameters is ${r}_{satellite}\sim~2$, and we attribute this to the fact that the placement of the `T' and `I' bars in the junction is not symmetric, causing a `bias' in favor of one of the two opposite paths. It is also interesting to note that when a droplet that enters the T-junction is below the threshold diameter limit ${d}_{min}$, it is diverted towards the path that has the `bias'. The performance of the T-junction for different arrangements of `T' and `I' bars and for higher frequencies $f$ as well as the derivation of a scaling law explaining this breakup will be saved for future efforts.
	
    
\subsection*{Study of droplet-to-droplet repulsions}

	Complementary to droplet break-up, we investigated the droplet-to-droplet repulsion . As previously established, two non-magnetic objects immersed in magnetic media are subject to a magnetostatic energy term\cite{toussaint2004:in}:
	
\begin{equation}
U=\frac{{\mu}_{f}}{4\pi}{\sigma}^{2}\Big(\frac{1-3{cos}^{2}\theta}{{r}^{3}})
\label{Interaction_energy}
\end{equation}	
	
 Based on this theory, a droplet-to-droplet repulsion force results from the minimization of this energy term from equation (\ref{Interaction_energy}). In fact, we calculated that the force ${F}_{repulsion}=-\vec{\bigtriangledown}U)$ is of the order of ${F}_{repulsion}\approx 50 \mu N$ therefore comparable a force from a polarized bar. Therefore, two droplets that come in proximity, driven by a magnetic force from a common magnetic pole will be subject to an additional repulsion force that may prevent their merging.  To investigate the antagonism between these two forces,  we reversed the direction of the ${\bf B}_{i}$ field in the junction for break-up (Fig. \ref {Figure_5}) and observed that we could not merge the satellite droplets into a single droplet. We also tested the droplet-to-droplet repulsion in specially designed geometry featuring a double bar geometry (\ref {Figure_6}a) to increase the magnetic force pulling the two droplets to the same location (\ref {Figure_6}b) but observed that the repulsive force between the droplets prevented their merging (\ref {Figure_6}d and supplementary video 6. Merging the droplets thus requires a different design where either the magnetic force of the bar is augmented by an additional local stimuli, for example a local magnetic field or a geometric feature that facilitates coalescence.

\section{Conclusions}

We have demonstrated magnetophoretic control of water droplets in bulk hydrocarbon oil-based ferrofluid. Even though the water droplets are essentially non-magnetic (water is weakly diamagnetic), they are subject to a magnetic force that is a result of their magnetic surrounding fluid. PIV analysis shows that the bulk ferrofluid is stationary unless there is a water in proximity. The magnetophoretic control of the water droplets is synchronous because it is based on a single magnetic field ${\bf B}_{i}$ that sets all droplets into motion. \revY{In addition, we have demonstrated a proof-of-principle droplet generator, designed tracks of bars for breaking-up droplets and showed repulsion forces in pairs of adjacent droplets.}

Extending the system towards direct applications would, however, require several investigations that we will save for future efforts. First, for the application domain it would be necessary to include external, `on-demand' user interference with the system which would require embedded electrodes placed beneath the permalloy bars. These electrodes could be actuated with electric currents to produce local magnetic fields that would selectively alter the behavior of the `T' and `I' bars. Second, in order to accommodate biology or chemical experiments with smaller droplets of diameters of the order of $d=10-50~\mu m$, it would be essential to scale down the system. The latter could be accomplished with standard microfabrications techniques, including but not limited to electroplatting \cite{barako_2015}. Third, the biocompatibility of the system would need to be checked; a water droplet is an ideal system for biological and chemical experiments \cite{teh2008:drop,schneider2013:pote} but the presence of the ferric bulk fluid could potentially result in iron anions (Fe+2 etc etc) leaching into the water droplet, as general crosstalk of components may occur \cite{chen2012cindytang}, thus depriving the droplets of its ability to act as a carrier of a number of experiments. Fourth, even if the biocompatibility can be guaranteed it should be necessary to develop de-emulsifying techniques to extract the bulk ferrofluid for subsequent steps in the analysis. Fifth, technical considerations need to be made based on recent advances with respect to improving the surface properties of the coating of the working surfaces to make them superoleophobic \cite{feng2006:de, tuteja2007:op,liu2009:bi,xue2012:re} as a means to increase the speed/frequency of the system. 	

\section{Acknowledgements}

We would like to acknowledge all members of Prakash Lab for useful discussions. G.K. was supported by the Onassis Foundation and the A.G. Leventis Foundation. A.B. was supported by funds from \'Ecole Polytechnique. M.P. was supported by the Pew Foundation, the Moore Foundation, the Keck Foundation, a Terman Fellowship, a NSF Career Award and a NIH Award. 






\bibliography{library} 
\bibliographystyle{rsc} 

\end{document}